\documentclass[11pt]{article}

\usepackage[hide]{todo}

\usepackage{psfrag}

\usepackage{amsthm}

\usepackage{pstool}

\usepackage{appendix} 

\usepackage{makeidx}

\usepackage{cite}

\usepackage[normalem]{ulem}
\usepackage{tensor}
\usepackage{emptypage}
\usepackage{slashed}
\usepackage{xcolor}
\usepackage{graphicx}  
\usepackage{psfrag}
\usepackage{epsf}
\usepackage{epic}
\usepackage{eepic}
\usepackage{epsfig}
\usepackage{wrapfig}
\usepackage{a4}
\usepackage{amssymb,latexsym}
\usepackage{amsfonts}
\usepackage[mathscr]{eucal}
\usepackage[active]{srcltx}
\usepackage{url}
\usepackage{fancybox}
\usepackage{mhequ}
\usepackage[g]{esvect}
\usepackage{epstopdf}

\usepackage{wasysym}

\usepackage{multirow}

\usepackage{tikz}  \usetikzlibrary{patterns}

\usepackage{bbm}
\usepackage{slantsc}
\usepackage{slashed}

\usepackage{marginnote}

\newcommand{\red}[1]{{\color{red}#1}}

\def\ben{\begin{equation}}
\def\een{\end{equation}}
\def\bena{\begin{eqnarray}}
\def\eena{\end{eqnarray}}

\def\f(#1/#2){\frac{#1}{#2}}
\def\Frac(#1/#2){\left(\frac{#1}{#2}\right)}
\def\chris(#1-#2-#3){{\mit \Gamma}^{#1}{}_{{#2}{#3}} }
\def\tilchris(#1-#2-#3){\tilde{{\mit \Gamma}}^{#1}{}_{{#2}{#3}}}
\def\hatchris(#1-#2-#3){\hat{{\mit \Gamma}}^{#1}{}_{{#2}{#3}}}

{\catcode `\@=11 \global\let\AddToReset=\@addtoreset}
\AddToReset{figure}{section}

\DeclareFontFamily{OT1}{rsfs}{}
\DeclareFontShape{OT1}{rsfs}{m}{n}{ <-7> rsfs5 <7-10> rsfs7 <10-> rsfs10}{}
\DeclareMathAlphabet{\mycal}{OT1}{rsfs}{m}{n}

{\catcode `\@=11 \global\let\AddToReset=\@addtoreset}
\AddToReset{equation}{section}

\newcounter{mnotecount}[section]

\renewcommand{\themnotecount}{\thesection.\arabic{mnotecount}}

\newcommand{\mnote}[1]
{\protect{\stepcounter{mnotecount}}$^{\mbox{\footnotesize
$
\bullet$\themnotecount}}$ \marginpar{
\raggedright\tiny\em
$\!\!\!\!\!\!\,\bullet$\themnotecount: #1} }

\definecolor{HP}{rgb}{1,0.09,0.58}

\newcommand{\eqref}[1]{\eq{#1}}

\newcommand{\hs}{\cH_{\mbox{\scriptsize sing}}}

\newcommand{\beadl}[1]{\begin{deqarr}\label{#1}}
\newcommand{\eeadl}[1]{\arrlabel{#1}\end{deqarr}}%

\def\nz{\ifmmode {I\hskip -3pt N} \else {\hbox {$I\hskip -3pt N$}}\fi}
\def\zz{\ifmmode {Z\hskip -4.8pt Z} \else
       {\hbox {$Z\hskip -4.8pt Z$}}\fi}
\def\qz{\ifmmode {Q\hskip -5.0pt\vrule height6.0pt depth 0pt
       \hskip 6pt} \else {\hbox
       {$Q\hskip -5.0pt\vrule height6.0pt depth 0pt\hskip 6pt$}}\fi}
\def\rz{\ifmmode {I\hskip -3pt R} \else {\hbox {$I\hskip -3pt R$}}\fi}
\def\cz{\ifmmode {C\hskip -4.8pt\vrule height5.8pt\hskip 6.3pt} \else
       {\hbox {$C\hskip -4.8pt\vrule height5.8pt\hskip 6.3pt$}}\fi}
\def\au{{\setbox0=\hbox{\lower1.36775ex\hbox{''}\kern-.05em}\dp0=.36775ex\hs
kip0pt\box0}}
\def\ao{{}\kern-.10em\hbox{``}}

\newcommand\Gregbeq{\begin{eqnarray}}
\newcommand\Gregeeq{\end{eqnarray}}

\newcommand{\scri}{{\mycal I}}%
\newcommand{\scripm}{\scri^{\pm}}%
\def\cH{{\cal H}}

\def\h1{{\hat 1}}
\def\h2{{\hat 2}}

\def\3f{\frac{3}{2}}

\def\th{{\tilde h}}

\newcommand{\oversetty}[2]{%
\mathop{#2}\limits^{\vbox to -.1ex{%
\kern -1.5ex\hbox{$\scriptstyle #1$}\vss}}}

\newcommand{\jlcax}[1]{}
\newcommand{\eean}{\nonumber\end{eqnarray}}

\newcommand{\kk}[1]{}

\newcommand{\beq}{\begin{equation}}

\newcommand{\rgc}[1]{}

\newcommand{\FS}       
                  {F}

\newcommand{\HS} 
       {H_{\mbox{\scriptsize volume}}}

{\ptc{this should be removed in the oberwolfach version}}%

\newcommand{\eel}[1]{\label{#1}\end{equation}}
\newcommand{\eeal}[1]{\label{#1}\end{eqnarray}}
\newcommand{\C}{{\mathbb C}}
\newcommand{\bed}{\begin{deqarr}}
\newcommand{\eed}{\end{deqarr}}
\newcommand{\bedl}[1]{\begin{deqarr}\label{#1}}
\newcommand{\eedl}[2]{\arrlabel{#1}\label{#2}\end{deqarr}}

\newcommand{\mcO}{{\mycal O}}

\newcommand{\bel}[1]{\begin{equation}\label{#1}}
\newcommand{\bea}{\begin{eqnarray}}
\newcommand{\bean}{\begin{eqnarray}\nonumber}
\newcommand{\beal}[1]{\begin{eqnarray}\label{#1}}
\newcommand{\eea}{\end{eqnarray}}

\newcommand{\be}{\begin{equation}}
\newcommand{\eeq}{\end{equation}}
\newcommand{\ee}{\end{equation}}
\newcommand{\beqa}{\begin{eqnarray}}
\newcommand{\eeqa}{\end{eqnarray}}
\newcommand{\beqan}{\begin{eqnarray*}}
\newcommand{\eeqan}{\end{eqnarray*}}
\newcommand{\ba}{\begin{array}}
\newcommand{\ea}{\end{array}}

\newcommand{\warn}[1]
{\protect{\stepcounter{mnotecount}}$^{\mbox{\footnotesize
$
\bullet$\themnotecount}}$ \marginpar{
\raggedright\tiny\em
$\!\!\!\!\!\!\,\bullet$\themnotecount: {\bf Warning:} #1} }

\newcommand{\R}{\mathbb{R}}

\newcommand{\N}{\mathbb N}

\newcommand{\eq}[1]{(\ref{#1})}

\newcommand{\ptc}[1]{\mnote{{\bf ptc:}#1}}

\newcommand{\beqar}{\begin{deqarr}}
\newcommand{\eeqar}{\end{deqarr}}

\newcommand{\beaa}{\begin{eqnarray*}}
\newcommand{\eeaa}{\end{eqnarray*}}

\newcommand{\bethm}{\begin{theorem}}
\newcommand{\et}{\end{theorem}}
\newcommand{\bl}{\begin{Lemma}}

\newcommand{\gtilde}{\red{\tilde g}}

\newcommand{\rMP}{\red{\hat r}}
\newcommand{\xMP}{\red{\hat x}}
\newcommand{\muMPhalf}{\red{ m}}
\newcommand{\muMP}{\red{2 \muMPhalf}}

\newcommand{\rstar}{\red{r_*}}
\newcommand{\rstarsquared}{\red{r_*^2}} 

\DeclareFontFamily{OT1}{rsfs}{}
\DeclareFontShape{OT1}{rsfs}{m}{n}{ <-7> rsfs5 <7-10> rsfs7 <10-> rsfs10}{}
\DeclareMathAlphabet{\mycal}{OT1}{rsfs}{m}{n}

{\catcode `\@=11 \global\let\AddToReset=\@addtoreset}
\AddToReset{equation}{section}

\renewcommand{\red}[1]{#1}

\title{Asymptotic flatness in higher dimensions}
\author{
  Peter Cameron\footnote{pjc96@cam.ac.uk}\\
  \small{Department of Applied Mathematics and Theoretical Physics,}\\
\small{University of Cambridge, Wilberforce Road, Cambridge CB3 0WA, UK.}
  \and
  Piotr T.\ Chru\'{s}ciel\footnote{piotr.chrusciel@univie.ac.at}\\
  \small{Faculty of Physics, University of Vienna,}\\
\small{Boltzmanngasse 5, A 1090 Vienna, Austria.}
}
\date{4 March 2022}
\usepackage{geometry}
\begin{document}

\maketitle

\abstract{We show that $(n+1)$-dimensional Myers-Perry metrics, $n\ge 4$, have a conformal completion at spacelike infinity of $C^{n-3,1}$ differentiability class, \red{and that the result is optimal in even spacetime dimensions. The associated asymptotic symmetries are presented.}}

\section{Introduction}
 \label{s4XI21.1}

One of the classical questions in general relativity is that of the behaviour of the gravitational field when receding to infinity in spacelike directions. This has been studied in detail in many works in spacetime dimension four, see e.g.~\cite{Penrose:PRL63,AshtekarHansen,BeigSimon2,Friedrich:i0,AshtekarRomano,BeigSchmidt,ChGeroch,ChAH,ChBeig3,Gerochizero,Hansen}.
In particular, as is well known,  Minkowski spacetime has a conformal completion that includes a point at spacelike infinity, called $\red{i_0}$, such that the conformally extended metric is smooth near this point. It is also well known that the usual extension of the Minkowskian construction to four-dimensional Schwarzschild spacetime leads to a conformal completion with a metric which is continuous~\cite{ChAH,AshtekarHansen,BeigSchmidt}, but no such extension which is $C^1$ is known.
Now, in standard examples of higher dimensional spacetimes the metric decays faster to the Minkowski metric at spacelike infinity, and therefore one expects that such constructions will lead to metrics with better differentiability properties at $\red{i_0}$ in higher dimensions. The question then arises about the precise degree of differentiability that one can achieve.

Now, the conformal method  of constructing solutions of the general relativistic constraint equations shows that the decay of the metric towards the flat one, when receding in spacelike directions, can be arbitrarily slow. Therefore nothing  striking can be said in this generality. But the study of asymptotics of initial data sets, and thus of the associated spacetimes, has been given new life by the work of Corvino and Schoen~\cite{CorvinoSchoen2} (compare~\cite{ChDelay}), who show that rather general initial data sets can be deformed at large distance to members of any family of metrics which contains a complete set of asymptotic charges.
A particularly useful such family is provided by the Myers-Perry metrics, to which boosts have been applied. The glued metric is then stationary near $\red{i_0}$, which gives one useful information concerning the resulting evolution.  This leads one naturally to raise the question of conformal properties near $\red{i_0}$ of spacetimes within this family.

The object of this note is to show that $(n+1)$-dimensional Myers-Perry metrics admit a conformal completion which includes a point $\red{i_0}$ at spatial infinity so that past and future null infinities $\scripm$ form the null cone emanating from this point, with a metric which is of $C^{n-3,1}$ differentiability class near this point. Thus, all derivatives of order less than $n-3$ extend continuously to $\red{i_0}$ and $\scri$, while the derivatives of the metric of order $n-3$ remain bounded as $\red{i_0}$ is approached, but do not necessarily extend to functions which are continuous at $\red{i_0}$.

We conclude  that the conformal structure at spacelike and null infinity is quite similar to that of Minkowski spacetime, except for a loss of differentiability at $\red{i_0}$ which is less severe than in the $(3+1)$-dimensional case. We also show that the result is optimal in \emph{even} spacetime dimensions in the following sense: no $C^{n-2}$ completions exist unless the total mass vanishes.

Recall, now, that the Riemannian metric obtained from quotienting the space-time metric of stationary, odd-dimensional, electrovacuum metrics  by the stationary isometries  admits a conformal completion  at a point at infinity which is real-analytic~\cite{ChBeig3}. But this result does not translate in any obvious way to the associated spacetime metric. And we do not know whether or not our result is optimal in such dimensions either.

Our observation concerning the differentiability of the conformally rescaled metric has immediate consequences concerning causality near $\red{i_0}$.
Recall that the usual results of causality theory, except perhaps the ones explicitly involving geodesics, apply for metrics which are Lipschitz continuous~\cite{ChGrant,MinguzziLRR}, and hence in dimensions $n\ge 4$ for metrics with the Myers-Perry asymptotics by our results here. Some care has to be taken in the borderline case $n=4$, since not all statements involving geodesics remain true for $C^{1,1}$ metrics. However, when  $n>4$ the full standard causality theory, including arguments involving geodesics, applies.

 Clearly the case $n=3$ requires special attention, since the conformal metric is only H\"older continuous at $\red{i_0}$ as constructed here or in \cite{AshtekarHansen,ChAH}. Recall that a function $f$ is H\"older continuous if there exist constants $C\ge0$, $\alpha>0$ such that $|f(x)-f(y)|\le C|x-y|^\alpha$ for all $x$ and $y$ in the domain of $f$. If $f$ and all its partial derivatives up to and including order $k$ are H\"older continuous with exponent $\alpha$, then we write $f\in C^{k,\alpha}$. It is not obvious whether or not the constructions can be improved to yield a conformally rescaled metric which is Lipschitz-continuous at $\red{i_0}$. However, the conformal metric is smooth everywhere except at $\red{i_0}$. It is then easily seen that  the causal bubbles of~\cite{ChGrant} do not occur. So, as emphasised in that last reference, standard causality theory, except possibly some arguments involving geodesics, applies also in this case.

This note is organised as follows: We start  with a short discussion of three dimensional spacetimes in Section~\ref{s22XII21.1}. We continue  in Section~\ref{s31X32.2} with the somewhat simpler case of the Schwarzschild metric in all spacetime dimensions $n+1\ge 4$.
The
Myers-Perry metrics are analysed in Section~\ref{s31X21.1}. We show in Section~\ref{s16XII21.1} that the ADM mass provides an obstruction to $C^{n-2}$-differentiability in even spacetime dimensions. Asymptotic symmetries of metrics with $C^{n-3,1}$-conformal completions are discussed in Section~\ref{s12XII21.1}.

\section{Three dimensional spacetimes}
 \label{s22XII21.1}

Three-dimensional spacetimes are of course special. The requirement that the metric be empty at large spacelike distances imposes flatness. If one assumes the existence of a well-behaved spacelike hypersurface, the space-geometry at large distances is that of a flat cone with a deficit angle. The deficit angle can \emph{a priori} have either sign, but positivity of matter energy implies positivity of this angle (cf., e.g.~\cite[Section 1.1.1]{ChEnergy}). Then
the spacetime metric is flat in the Minkowskian domain of dependence of the angular sector
\begin{equation}
 \label{22XII21.1}
 \{ z\in \C, \ |z| \ge R, \arg z \in [0, \alpha]
  \} \subset \{t=0\}
  \,,
\end{equation}
with $\alpha \in (0,2\pi)$ except if there is no matter, and with $R >0$, where  the boundary $\arg z =0$ should be identified with the boundary $\arg z = \alpha$.
The Lorentzian Kelvin transformation \eqref{30IX21.5} with $r^*=|z|$ leads, after addition of a point $i_0=\{ t=1/|z| = 0\}$, to a conformal completion where the conformally rescaled metric has a deficit angle of $2\pi - \alpha$ at $i_0$ on the spacelike hyperplane $\{\tau=0\}$, and the null conformal  boundary at infinity, $\scri$, is the light-cone emanating from $i_0$ with the same deficit angle.

\section{The Schwarzschild metric}
 \label{s31X32.2}

Let $d\Omega^2_{n-1}$ denote the unit round metric on $S^{n-1}$. We rewrite the $(n+1)$-dimensional Schwarzschild metric as
\begin{eqnarray}
 g
  &=&
    - (1-\frac {2m}{r^{n-2}} ) dt^2 + \frac {dr^2 } {1-\frac {2m}{r^{n-2}} }
 +r^2 d\Omega^2_{n-1}
 \nonumber
\\
  &=&
     (1-\frac {2m}{r^{n-2}} )
    \big(
      - dt^2 +
       \underbrace{ \frac { dr^2} {(1-\frac {2m}{r^{n-2}})^2 }
}_{=:d\rstarsquared }
  \big)
    +r^2 d\Omega^2_{n-1}
 \nonumber
\\
  &=&
     (1-\frac {2m}{r^{n-2}} )
    \Big(
      - dt^2 +
      {d\rstarsquared }
      + \rstarsquared d\Omega^2_{n-1}
      \nonumber
\\
 &&
 \phantom{  (1-\frac {2m}{r^{n-2}} )
    \Big(}
    + \big( \frac {r^2} {1-\frac {2m}{r^{n-2}} } -\rstarsquared  \big) d\Omega^2_{n-1}
  \Big)
   \,.
 \label{30IX21.1}
\end{eqnarray}
When ignoring the last correction term, and up to a conformal factor, this is the Minkowski metric with a radial coordinate
\begin{equation}\label{30IX21.2}
  \rstar  = \int_{0}^{r} \frac{dr'}{1-\frac {2m}{r'^{n-2}} }
  =
  \left\{
    \begin{array}{ll}
      r + 2m \ln \big( \frac{r}{2m}-1\big), & \hbox{$n=3$;} \\
      r +\frac{2 m}{3-n}r^{3-n}
+\frac{4 m^2}{5-2
   n}r^{5-2 n}
 +\frac{8 m^3}{7-3 n}r^{7-3 n}+ \ldots , & \hbox{$n\ge 4$.}
    \end{array}
  \right.
\end{equation}
 Now, the Minkowski part of the metric can be conformally completed by adding a point at infinity $\red{i_0}$: letting
\begin{equation}\label{30IX21.5}
 \tau = \frac{t}{\rstarsquared  - t^2}
 \,,
 \qquad
 \rho= \frac{\rstar }{\rstarsquared  - t^2}
\,,
\end{equation}
we have
$$
 - dt^2 +
      {d\rstarsquared }
      + \rstarsquared d\Omega^2_{n-1} =
\frac{1}{(\rho^2 - \tau^2)^2}
\big(
 - d\tau^2 +
      {d\rho ^2}
      + \rho^2d\Omega^2_{n-1}
\big)
\,,
$$
with the point $\tau=0=\rho$ representing $\red{i_0}$.
This allows us to rewrite \eqref{30IX21.1} as

\begin{eqnarray}
 g \equiv \Omega^{-2} \tilde g
  &=&
     \underbrace{\frac{1-\frac {2m}{r^{n-2} }}{(\rho^2 - \tau^2)^2}
}_{=:\Omega^{-2}}
    \Big(
      - d\tau^2 +
      {d\rho ^2}
      + \rho^2d\Omega^2_{n-1}
      \nonumber
\\
 &&
 \phantom{  \frac {2m}{r^{n-2}} )}
    + \underbrace{
\underbrace{
 (\rho^2 - \tau^2)^2
\big( \frac {r^2} {1-\frac {2m}{r^{n-2}} } -\rstarsquared  \big)
 }_{=:\psi} d\Omega^2_{n-1}
}_{=:\delta \tilde{g}}
  \Big)
   \,.
 \label{30IX21.7}
\end{eqnarray}
The question arises of the regularity  at $\red{i_0}$ of
the conformal factor $\Omega$,
 of
the function $\psi$ and of the  correction term $\delta \tilde{g}$.
For this we will need to invert \eqref{30IX21.5}:
\begin{equation}\label{30IX21.6}
 t= \frac{\tau}{\rho^2 - \tau^2}
 \,,
 \qquad
 \rstar  = \frac{\rho }{\rho^2 - \tau^2}
\,.
\end{equation}

We consider first the somewhat simpler case $n\ge4$, since then $r$ asymptotes to $\rstar $, and there are no log terms. One then finds that there exist smooth functions $f\,,\, f_*\,,\, h\,,\, h_*:\R\to \R$, with $f(0)=f_*(0)=h(0)=h_*(0)= 0$, such that
for large $r$, and assuming that all coordinates as well as the mass are unitless, we have
%
\begin{eqnarray}
  \rstar
 &=& r
 \left(1 - \frac{2m}{(n-3)r^{n-2}} \big(1+ f( r^{2-n})
 \big)
  \right)
\nonumber
\,,
 \\
 \nonumber
  r  &=& \rstar
 \left(1 + \frac{2m}{(n-3)\rstar ^{n-2}}
    \big(1+ f_*( \rstar ^{2-n})
 \big)
  \right)
\,,
 \\
  \frac {r^2} {1-\frac {2m}{r^{n-2}} } -\rstarsquared
 &=&  \frac{2(n-1)m}{(n-3)r^{n-4}}
    \big(1+ h( r ^{2-n})
 \big)
 \nonumber
 \\
 &=&  \frac{2(n-1)m}{(n-3)\rstar ^{n-4}}
    \big(1+ h_*( r _*^{2-n})
 \big)
 \nonumber
 \\
 &=&
  \frac{2(n-1)m}{(n-3)}
  {\left(  \rho \big(1 - \frac{\tau^2}{\rho^2 }\big)\right)}^{n-4}
      \left(1+ h_* \big( \big(\frac{\rho }{\rho^2 - \tau^2}\big)^{2-n}\big)
 \right)
 \nonumber
\,,
 \\
 \rho^{-2} \psi
 &=&
  \frac{2(n-1)m}{(n-3)} \rho^{n-2} \left( 1-\frac{  \tau^2}{\rho^2 }\right)^{n-2}
   \times
     \nonumber
\\
      &&
    \Big(1+
  h_* \big( \rho^{n-2} \big( 1-\frac{  \tau^2}{\rho^2 }\big)^{n-2}\big)
 \Big)
 \nonumber
\,,
\\
 \delta \tilde g  \equiv
 \psi  d\Omega^2_{n-1}
 &=&
 \psi \rho^{-2} ((dx^1)^2 + \ldots + (dx^n)^2 - d\rho^2 )
\nonumber
\\
 &=&
 \psi \rho^{-2}
 \big(
 (dx^1)^2 + \ldots + (dx^n)^2 -\rho^{-2}  (x^1 dx^1  + \ldots +  x^n dx^n)^2
 \big )
\,,
 \phantom{xxxxxx}
\label{1X21.11}
\end{eqnarray}
where $(x,y,z)$ are the usual Cartesian coordinates on $\R^3$, with $\rho^2 = x^2 + y^2 + z^2$.

Now, in the coordinates above the original Schwarzschild metric is defined only in the region $-\rho < \tau < \rho$. Hence it is convenient to extend  the function $\tau^2/\rho^2$ beyond $\tau=\pm \rho$ by any smooth function which is constant for, say, $|\tau| \ge 2 \rho$. Then  both  $\delta \tilde g$  and $\tilde g$, so extended, are continuous in a neighborhood of $\{\rho =0\}$  for all $n\ge 4$.

Next, the function $\lambda:=\psi \rho^{-2}$  appearing in \eqref{1X21.11}. has the property that  for all multi-indices $\alpha\in \N^{n+1}$ we have, for small $\rho$,

$$ \rho ^{\alpha_0+\ldots +\alpha_n} \partial_0 ^{\alpha_0 }
 \cdots
\partial_n ^{\alpha_n }
  \lambda=O(\rho^{n-2})
\,.
$$
A similar property holds for the functions $\lambda x^i x^j /\rho^2$:
$$ \rho ^{\alpha_0+\ldots +\alpha_n} \partial_0 ^{\alpha_0 }
 \cdots
\partial_n ^{\alpha_n }
 \left( \lambda
 \frac{ x^i x^j}{  \rho^2}
\right)
=O(\rho^{n-2})
\,.
$$
 It follows that $\lambda\in C^{n-3,1}$ for   $n\ge 4$, which shows that   $\tilde g$ is of differentiability class $C^{n-3,1}$ near $i_0$.
 
We note that, in even space dimensions, both the conformal factor $\Omega$ and the metric induced on the slices $\tau=0$, as constructed above, are smooth, even analytic. This is a special case of \cite{ChBeig3}, where analyticity at $i_0$ is established  on static slices for all metrics which are static and vacuum at large distances in even space dimensions larger than 6; analyticity is also established there in dimension four if one  assumes in addition the non-vanishing of the  ADM mass.

For completeness we apply the method above to the $n=3$ case, already analysed in~\cite[Appendix~C]{AshtekarHansen} from a somewhat different perspective, with two completions,  better behaved than the one here, constructed in \cite{ChAH}:
\begin{eqnarray}
  \rstar
 &=& r + 2m \ln (\frac{r}{2m} - 1)
\nonumber
\,,
 \\
 \nonumber
  r  &=& \rstar  - 2m \ln (\frac{\rstar }{2m})
 +  \frac{4 m^2 \left(\ln  \left(\frac{\rstar }{2
   m}\right)+1\right)}{\rstar }
 + O \big(\frac{\ln^2 \rstar }{\rstarsquared }\big)
\,,
 \\
  \frac {r^2} {1-\frac {2m}{r} } -\rstarsquared
 &=&  - 4 m \rstar \ln (\frac{\rstar }{2m}) + 2m \rstar + O ( \ln ^2(\frac{\rstar }{2m}) )
 \nonumber
\\
 &=&  -  \frac{4 m\rho }{\rho^2 - \tau^2}\ln (\frac{\rho }{\rho^2 - \tau^2}) + 2m \frac{ \rho }{\rho^2 - \tau^2}
 + O \big( \ln ^2\big(\frac{\rho }{\rho^2 - \tau^2}\big) \big)
 \nonumber
\,,
 \\
  \psi
 &=&   -   {4 m\rho }{(\rho^2 - \tau^2)}\ln (\frac{\rho }{\rho^2 - \tau^2}) + 2m  { \rho }(\rho^2 - \tau^2)
 \,,
 \nonumber
\\
 &&
 + O \big( (\rho^2 - \tau^2)^2 \ln ^2\big(\frac{\rho }{\rho^2 - \tau^2}\big) \big)
 \nonumber
\\
\rho^{-2}  \psi
 &=&       {4 m\rho }\big( 1 - \frac{\tau^2 }{\rho^2}
\big)
\big(
 \ln \big(1- \frac{\tau^2}{\rho^2 }\big)+\ln ( {\rho }) - 2m  { \rho }
\big)
 \nonumber
\\
 &&
 + O \big( \rho^{ 2}  \big( 1 - \frac{\tau^2 }{\rho^2}
\big) ^2 \ln ^2\big(\frac{\rho }{\rho^2 - \tau^2}\big) \big)
 \,,
 \label{1X21.1}
\\
 \delta \tilde g  \equiv
 \psi  d\Omega^2_{2}
 &=&
 \psi \rho^{-2} (dx^2 + dy^2 + dz^2 - d\rho^2 )
\,.
\end{eqnarray}
Note that for any $\alpha \in (0,1)$ the function $\big( 1 - \frac{\tau^2 }{\rho^2}
\big)
 \ln \big(1- \frac{\tau^2}{\rho^2 }\big)$ extends in $C^{0,\alpha}$ across the light-cone $\tau = \pm \rho$, but is not Lipschitz.
Similarly $\rho \ln \rho$ is $C^{0,\alpha}$ but not $C^{0,1}$. It follows that, in the coordinates above, $\tilde g$ can be extended across $\tau = \pm \rho$ to a metric satisfying $\tilde g \in C^{0,\alpha}$ for any $\alpha \in (0,1)$, but $\tilde g \not \in C^{0,1}$.

An analysis similar to the above shows that the conformal factor in \eqref{30IX21.7}, namely
\begin{equation}\label{1XII21.1}
  \Omega = \frac{\rho^2 - \tau^2}{\sqrt{1-\frac {2m}{r}}}
  \,,
\end{equation}
consists of a smooth function $\rho^2-\tau^2$ multiplied by a function which is $C^{1,\alpha}$ across the light-cone for all $\alpha\in[0,1)$, but is not $C^{1,1}$ there.

We can remove this problematic function by absorbing a factor of $1-\frac {2m}{r }$ into $\tilde g$. This does not change the differentiability properties of $\tilde g$,  and the resulting conformal factor,  $\rho^2 - \tau^2$, is a smooth function on the conformally completed manifold.

\section{Myers-Perry metrics}
 \label{s31X21.1}


In Kerr-Schild-type coordinates the Myers-Perry metrics take the form~\cite{myersperry}
\begin{equation}\label{31X21.1}
  g_{\mu\nu} = \eta_{\mu\nu} + h k_\mu k_\nu
  \,,
\end{equation}
where $k_\mu$ is a null vector for the $(n+1)$-dimensional Minkowski metric $\eta$, and hence also for the full metric:
\begin{equation}\label{31X21.2}
  g^{\mu\nu} = \eta^{\mu\nu} - h k^\mu k^\nu
  \,,
 \quad
 \eta^{\mu\nu}k_\mu k_\nu=g^{\mu\nu}k_\mu k_\nu=0
  \,,
 \quad
 k^\nu = \eta^{\mu\nu}k_\mu =g^{\mu\nu}k_\mu
  \,
\end{equation}
and $h$ is a function to be defined later.

In the calculations to follow we assume $n\ge 3$, and note that $n=3$ is the Kerr metric in Kerr-Schild coordiantes.

In order to handle even and odd dimensions simultaneously, we introduce a parameter
$$
 \epsilon = \left\{
                      \begin{array}{ll}
                        0, & \hbox{$n=2d$ is even;} \\
                        1, & \hbox{$n=2d+1$ is odd,}
                      \end{array}
                    \right.
$$

and write
\begin{eqnarray}
 (x^\mu)  &\equiv& (x^0,x^i) = (t,x_1,y_1,\ldots,x_d,y_d,\epsilon z)
    \nonumber
     \\
    &=&  \big(t, r_1 \cos(\varphi_1),  r_1 \sin (\varphi_1), \ldots,r_d \cos(\varphi_d),  r_d \sin (\varphi_d),r_{d+1}
  \big)
  \,,
   \phantom{XXX}
\end{eqnarray}
with the Minkowski metric  taking the form
$$
\eta = -dt^2 + \sum_{i=1}^{d} (dx_i^2+dy_i^2) + \epsilon dz^2= -dt^2 + \sum_{i=1}^{d} (dr_i^2+r_i^2 d\varphi_i^2) + dr_{d+1}^2
\,.
$$
Note that with the definitions above, $r_{d+1}\equiv 0$ in even space dimensions $n=2d$, so $ dr_{d+1} \equiv 0$.
\\
\\
In these coordinates, the vector field $k$ is
\begin{eqnarray}
  k_\alpha  dx^\alpha &=&
   dt + \sum_{i=1}^{d}  \frac{\rMP (x_i dx_i + y_i dy_i) +a_i (x_i dy_i - y_i dx_i) }{\rMP ^2+a_i^2}+\epsilon\frac{zdz}{\rMP}
    \nonumber
\\ &=&
   dt + \sum_{i=1}^{d+1}  \frac{\rMP r_i dr_i + a_i r_i^2   d\varphi_i  }{\rMP ^2+a_i^2}
   \label{31X21.6}
\end{eqnarray}
where $a_i\in \R$ are constants and we have set
$$
a_{d+1}=0.
$$
We also define the function $\rMP $ implicitly by the condition that $k$ is null:
%
\begin{eqnarray}
  \sum_{i=1}^{d}\frac{ x_i^2+y_i^2 }{\rMP ^2+a_i^2}+\epsilon\frac{z^2}{\rMP^2} = 1
  \label{31X21.7}
  \quad
  \Longleftrightarrow
  \quad
  \sum_{i=1}^{d+1}\frac{ r_i^2 }{\rMP ^2+a_i^2} = 1
  \,.
\end{eqnarray}
The function $h$ in \eqref{31X21.2} is defined to be
\begin{eqnarray}
  h &=& \frac{\muMP  \rMP^{2-\epsilon} }{\Pi F}
\end{eqnarray}
where
\begin{eqnarray}
  F &=& 1 - \sum_{i=1}^{d}\frac{a_i^2(x_i^2+y_i^2)}{(\rMP ^2+a_i^2)^2}
  \,,
\\
  \Pi &=&  \prod_{i=1}^{d} (\rMP ^2+a_i^2)
  \,.
\end{eqnarray}
Letting
$$
 r^2 := \sum_{i=1}^{d+1} r_i^2
  \equiv \sum_{i=1}^{d} (  x_i^2+y_i^2)+
   \epsilon z^2
 \,,
$$
one finds the following asymptotic expansions for large $r$
%
\begin{eqnarray}\label{31X21.8}
  \rMP^2
   & = &
    r^2  - \sum_{i=1}^{d} \frac{a_i^2r_i^2}{r^2} + O(r^{-2})
    =
    r^2\big(1 + O(r^{-2})
  \big)
    \,,
\\
 F & = & 1 + O(r^{-2})
    \,,
\\
 \Pi & = & r^{n-\epsilon}\big(1 + O(r^{-2})
  \big)
    \,,
\\
 h & = &  \frac{\muMP }{r^{  n-2 }}\big(1 + O(r^{-2})
  \big)
\\
  g &=&  \eta + \frac{\muMP }{r^{  n-2 }}\Big(
    dt^2
  +
   2dt\sum_{i=1}^{d+1}  \frac{ rr_i dr_i +a_i r_i^2 d\varphi_i  }{r^2}
  \nonumber
\\
 &&
  +
     \sum_{i,j=1}^{d+1}  \frac{
     \big(
      rr_i dr_i +a_i r_i^2 d\varphi_i
      \big)
     \big(
      rr_j dr_j +a_j r_j^2 d\varphi_j
      \big)
      }{r^4 }
  \nonumber
\\
   &&
  + O(r^{-2}) dx^\mu    dx^\nu
  \Big)
  \,.
\end{eqnarray}

We see from the last equation that the metric is asymptotically flat at spatial infinity, with well defined energy, but the mixed time-space metric functions decay slower by one power of $r$ than usually expected. This can be remedied by a  fine-tuning of the coordinates, but will not be necessary for our purposes.

We can carry out a variant of the calculations of Section~\ref{s31X32.2} as follows: We write the metric as  in \eqref{31X21.2} and perform a Kelvin transformation as in \eqref{30IX21.5}, with $\rstar$ there replaced by $r$,
\begin{equation}\label{30IX21.6435}
 t= \frac{\tau}{\rho^2 - \tau^2}
 \,,
 \qquad
 r = \frac{\rho }{\rho^2 - \tau^2}
\,,
\end{equation}
to obtain:
\begin{eqnarray}
  g &=& \eta +  h\,  k \otimes k
     \nonumber
\\
 &=&
   \underbrace{\frac{1}{(\rho^2-\tau^2)^2 }
    }_{=:\Omega^{-2}}
    \Big(
    \underbrace{
     - d\tau^2 + d\rho^2 + \rho^2 d\Omega^2_{n-1}
      }_{\eta}
     +  (\rho^2-\tau^2)^2
     h\,  k \otimes k
     \Big)
   \,.
\end{eqnarray}
To understand the properties of $\Omega^2 g$ at
\begin{equation}\label{1X21.12}
 \red{i_0}:=\{\rho=\tau=0\}
\end{equation}
we need to analyse the   function $h$ and the covector field
\begin{equation}
     \tilde  k := (\rho^2-\tau^2) k
     \,.
\end{equation}
For this we return to  \eqref{31X21.7}, which after setting
$$
 r_i := r \, m_i \, (i=1,...,d+1)
 \,,
 \qquad
 x:= 1/r\,,
 \qquad\xMP:= 1/\rMP
$$
(thus  $\sum_{i=1}^{d+1} m_i^2 =1$),   can be rewritten as
\begin{equation}\label{31X21.72}
   \xMP \underbrace{
    \sqrt{
     \sum_{i=1}^{d+1}\frac{ m_i^2 }{1+a_i^2\xMP^2 }
     }
    }_{=: f_1(\xMP^2,m_i)} = x
  \,.
\end{equation}

The function $0< f_1 \le 1$ is real analytic, with $f_1(0,m_i)=1$. One easily checks that $\xMP\mapsto \xMP f_1(\xMP^2,m_i)$ is strictly increasing  so that, for any given collection $(m_i)$, by the (analytic) inverse function theorem for real-valued functions defined on subsets of $\R$   there exists a real-analytic function $s\mapsto f_2(s,m_i)>0$ solving \eqref{31X21.72}:
\begin{equation}\label{31X21.73}
   \xMP  =  x f_2(x^2,m_i)\,, \ \mbox{with} \ f_2(0,m_i)=1
  \,.
\end{equation}
In terms of $(\tau,\rho)$ coordinates this reads
\begin{equation}\label{31X21.73}
   \rMP = \frac{
    1}{
       \rho \big( 1 - \frac{\tau^2}{\rho^2}
     \big) f_2\big(\rho^2(1 - \frac{\tau^2}{\rho^2})^2,m_i\big)}=:
     \frac{ f_3(\tau^2,\rho^2,m_i)}{ \rho \big( 1 - \frac{\tau^2}{\rho^2}
     \big)}
      \,, \ \mbox{with} \ f_3(0,0,m_i)=1
  \,.
\end{equation}
As in the Schwarzchild case we are only interested in the region $|\tau/\rho|\le 1$. It is therefore convenient to extend the function $1-\tau^2/\rho^2$ beyond this region to a smooth function which is constant for $|\tau/\rho|\ge 2$. Then the function $f_3$ is a smooth function of its arguments on the set
$$\mcO:=\{\tau^2+\rho^2<\delta\}\times\{(m_i)\in S^{d+\epsilon-1}\}
\,,
$$
for some $\delta>0$, where $S^n$ denotes the unit $n$-sphere for $n\ge1$.  It also follows that the functions $r^{-2}$, $\rMP^{-2}$, $r/\rMP$ and $\rMP/r$ are smooth on $\mcO$, with $r/\rMP|_{i_0}=1$.

One now checks that there exist functions $f_{\cdot}$, all smooth on $\mcO$, such that:
\begin{eqnarray}
 F & = & 1- \sum_{i=1}^{d}\frac{a_i^2r^2 m_i^2 }{(\rMP ^2+a_i^2)^2}
  = 1 - \rho^2 f_4
    \,,
\\
 \Pi^{-1} & = &  \prod_{i=1}^{d} (\rMP ^2+a_i^2)^{-1}
 = \rho ^{n-\epsilon}
  \big(1 - \frac{\tau^2}{\rho^2}\big)^n \big(1 + \rho^2f_5)
  \big)
    \,,
\\
 h & = & \frac{\muMP  \rMP ^{2-\epsilon}}{\Pi F}=
  \muMP \rho ^{n-2}
  \big(1 - \frac{\tau^2}{\rho^2}\big)^{n-2} \big(1 + \rho^2f_6)
  \big)
  \,,
\\
 \tilde k  &=&  \rho^2 \big(1-\frac{\tau^2}{\rho^2}\big)
 \big(
    dt +  \sum_{i=1}^{d}  \frac{ \rMP r_i dr_i +a_ir_i^2 d\varphi_i }{\rMP^2 + a_i^2}+\epsilon\frac{zdz}{\rMP}
    \big)
  \nonumber
\\
&=&  \rho^2 \big(1-\frac{\tau^2}{\rho^2}\big)
 \Big(
    dt + \sum_{i=1}^{d+1} \big(
       m_i d(r m_i)(1+\rho^2 f_{7,i})
         +a_i m_i^2 (1+\rho^2 f_{8,i}) d\varphi_i
    \big)
    \Big)
  \nonumber
\\
&=&
\rho^2 \big(1-\frac{\tau^2}{\rho^2}\big)
 \Big(
    \underbrace{d(t + r)}_{(d\tau-d\rho)(\rho-\tau)^{-2}} +  \underbrace{\sum_{i=1}^{d+1}\Big(r
        m_i d  m_i }_{=0}
  \nonumber
\\
& &
       +
       \underbrace{
        \rho^2 f_{7,i}  m_i d(r m_i) }
         _{
        \rho^3 f_{7,i}  m_i d m_i   /(\rho^2-\tau^2)
         +
        \rho^2 f_{7,i}  m_i^2  d r  }
         +a_i m_i^2 (1+\rho^2 f_{8,i}) d\varphi_i
    \Big)\Big)
  \nonumber
\\
&=&
 \frac{1+\frac{\tau}{\rho}}{1-\frac{\tau}{\rho}}
 (d\tau-d\rho)
    +
    \sum_{i=1}^{d+1} \Big(
        \rho^3  f_{7,i}  m_i d m_i
  \nonumber
\\
& &
    + \rho^2 \big(1-\frac{\tau^2}{\rho^2}\big)
 \big(
        \rho^2 f_{7,i}  m_i^2  d r
         +a_i m_i^2 (1+\rho^2 f_{8,i}) d\varphi_i
         \big)
    \Big)
  \,.
\end{eqnarray}
Using
\begin{equation}\label{1X21.51}
  dr =
    \rho^{-2}\big(1-\frac{\tau^2}{\rho^2}\big)^{-2}
  \big( \frac{2\tau}{\rho } d\tau -  \big(1+\frac{\tau^2}{\rho^2}\big) d\rho
  \big)
  \,,
\end{equation}
we can rewrite $\tilde k$ as
\begin{eqnarray}
 \tilde k
&=&
 \frac{1+\frac{\tau}{\rho}}{1-\frac{\tau}{\rho}}
 (d\tau-d\rho)
   +
   \sum_{i=1}^{d+1} \big(
        \rho^3 f_{7,i}  m_i d m_i
         +a_i \rho^2 m_i^2
         f_{9,i}
   d\varphi_i
    \big)
    \Big)
  \nonumber
\\
& & +
       \frac{1}{1-\frac{\tau^2}{\rho^2}}
          \sum_{i=1}^{d+1}    f_{7,i}  m_i^2
        \big( 2 \rho \tau d \tau
        -   \big(\rho^2 + \tau^2) d\rho
        \big)
  \,.
\end{eqnarray}
It follows that $h\,\tilde k \otimes \tilde k$ takes the form
\begin{eqnarray}
  h \, \tilde k \otimes \tilde k &=& \muMP  \rho^{n-2} \big(1-\frac{\tau^2}{\rho^2}\big)^{n-4} \psi_{\mu\nu} dy^\mu dy^\nu
  \,,
\end{eqnarray}
where $(y^\mu)\equiv (y^0,y^i):= (\tau, \rho m^i)$, and where the functions $\psi_{\mu\nu}$ satisfy, for all $\alpha\in \N^{n+1}$,
$$ \rho ^{\alpha_0+\ldots +\alpha_n} (\partial_{y^0}) ^{\alpha_0 }
 \cdots
(\partial_{y^n} )^{\alpha_n }
  \psi_{\mu\nu}=O(1)
\,.
$$
If $n\ge 4$  this leads, as in the Schwarzschild case,  to a conformally extended metric of differentiability class $C^{n-3,1}$.
When $n=3$ the behaviour of this completion at $i_0$  along spacelike directions is better than in the completion of Section~\ref{s31X32.2}, but is worse when the light cone is approached.

\section{The ADM mass as an obstruction to differentiability}
 \label{s16XII21.1}

So far we have transformed the metric to the form
\begin{equation}\label{1XII21.3}
  g = \frac{1}{(\rho^2-\tau^2)^2}\gtilde
  \,,
\end{equation}
with
\begin{equation}\label{1XII21.4}
  \gtilde_{\mu\nu} = \eta _{\mu\nu} + \th_{\mu\nu}
  \,,
  \quad
  \left\{
    \begin{array}{ll}
     \th_{\mu\nu}=O(\rho \ln \rho ), & \hbox{$n=3$,}
\\
     \th_{\mu\nu}=O(\rho^{n-2}), & \hbox{$n\ge 4$,}
    \end{array}
  \right.
\end{equation}
and  with $\rho \equiv |\vec y|$.
We want to show that our differentiability conditions obtained so far are optimal, in the following sense:

Suppose for contradiction that, in
 odd
space dimensions with $n\ge 3$,   there exists a coordinate system $(y^\mu)\equiv (\tau, \vec y)$ in which   $\th_{\mu\nu}=O(\rho^{n-2})$ and $\th_{\mu\nu}$ is of differentiability class $C^{n-2}$.
Consider the metric $\gamma:= g_{ij}dx^idx^j$ induced by $g$ on the zero level-set of $\tau$, where the coordinates $x^i$ are obtained by a Kelvin-inversion from the coordinates $y^i$:
\begin{equation}\label{1XII21.5}
  x^i = \frac{y^i}{\rho^2}
   \,.
\end{equation}
We find
\begin{equation}\label{1XII21.6}
  g_{ij} = \delta_{ij} + \th_{ij}
   -4 \frac{\th_{k(i}x^k x_{j)}}{r^2}
    + 4\frac{\th_{k \ell}x^k x^\ell x_i x_j}{r^4}
  \,,
\end{equation}
where for visual convenience we have set $x_i:= x^i$.
This leads to the following ADM integrand, up to a constant multiplicative factor and the measure on $S^{n-1}$:
\begin{eqnarray}
 \mu
 &  \equiv  &
  U_i \frac{x^i}{r} \times r^{n-1}
 : =
  \big(
   \frac{\partial g_{ij}}{\partial x^j}
  -  \frac{\partial g_{jj}}{\partial x^i}
  \big) \frac{x^i}{r} \times r^{n-1}
 \nonumber
\\
 &=&
 \Big(
    \frac{\partial \th_{ij}}{\partial x^j}
   -4 \frac{\frac{\partial \th_{k(i}}{\partial x_j}x^k x_{j)}}{r^2}
    + 4\frac{\frac{\partial \th_{k \ell}}{\partial x_j} x^k x^\ell x_i x_j}{r^4}
  -  \frac{\partial \th_{jj}}{\partial x^i}
  \nonumber
\\
 & & -4\th_{k(i} \partial_j\big(  \frac{x^k x_{j)}}{r^2}
 \big)
    + 4\th_{k \ell} \partial_j\big(\frac{x^k x^\ell x_i x_j}{r^4}
 \big)
   \Big) \frac{x^i}{r} \times r^{n-1}
  \,.
  \label{2XII21.1}
\end{eqnarray}
Suppose that $\th_{ij}$ is of differentiability class $C^{n-2}$. Then $ \th_{ij} $ is  a homogeneous polynomial of order $n-2$ in $\vec y$ up to terms $o(\rho^{n-2})$ and $
\partial \th_{ij}/\partial y^k$ is  a homogeneous polynomial of order $n-3$ in $\vec y$  up to terms $o(\rho^{n-3})$.
Using
$$
 \frac{\partial \th_{ij}}{\partial x^k}  = \frac{\partial \th_{ij}}{\partial y^\ell}
 \frac{(r^2 \delta^\ell_k - 2 x^\ell x^k)}{r^4}
 = \frac{\partial \th_{ij}}{\partial y^\ell} (\rho^2 \delta^l_k- y^\ell y^k)
\,,
$$
this implies that the right-hand side of \eqref{2XII21.1}
is a sum of homogeneous polynomials in $\vec x/r$  of odd orders, up to $o(1)$ terms.

When $n$ is odd this implies immediately that the ADM mass vanishes. We recall that the constructions in~\cite[Section~III]{ChAH} in spacetime dimension $n+1=4$ provide  completions of the Schwarzschild metric at $\red{i_0}$ \emph{either} without $\rho\log\rho$ terms  \emph{or} without the $(1-\tau^2/\rho^2)\log(1-\tau^2/\rho^2)$ terms. It is not known whether such terms can be removed altogether to obtain a $C^{0,1}$ metric.

While the above calculations apply for all $n\ge 3$, the vanishing of mass is not true if the spatial dimension, $n$, is even.
As pointed out at the end of Section~\ref{s31X32.2}, in even dimensions the conformal compactification of the Riemannian hypersurfaces of constant $t$, induced by the spacetime compactification of the Schwarzschild metric described there, leads to a real-analytic conformal metric with a real-analytic conformal factor.
%
%
%
%

\section{Asymptotic symmetries}
 \label{s12XII21.1}

Somewhat more generally, one might wish to consider the collection of metrics with asymptotic behavior captured by \eqref{1XII21.3} and \eqref{1XII21.4}.

A natural question  is then that of the set of coordinate transformations compatible with a conformal compactification of this form; compare~\cite{Prabhu}. A detailed analysis of this in four-dimensional spacetimes can be found in~\cite{ChAH}.

First, there are  Lorentz transformations in the coordinates $y^\mu$, which preserve both the conformal factor $\Omega$ and the asymptotic behavior of the metric.

Next,  there are coordinate transformations of the form
\begin{equation}\label{12XII21.11}
  y^\mu \mapsto y^\mu + \xi^\mu(\tau/\rho, y^i/\rho)\rho^{n-1}
  \,,
\end{equation}
with $\xi^\mu$ a smooth functions of its arguments. These generalise to higher dimensions the usual ``supertranslations'' at $\red{i_0}$. There is, however, a key difference: in four spacetime dimensions the Hamiltonian charges associated with the coordinate transformations \eqref{12XII21.11} are non-zero in general, while in higher dimensions the decay in  \eqref{12XII21.11} is too fast, leading to vanishing Hamiltonian generators.

We note that the four-dimensional   \emph{logarithmic ambiguities},
\begin{equation}\label{12XII21.11at}
  x^\mu \mapsto x^\mu + \eta_{\alpha\beta}(C^\mu  x^\alpha  - 2  C^\alpha x^\mu  ) x^\beta\ln |\vec x|
  \,,
\end{equation}
with constants $C^\mu$ (see~\cite[Theorem~2]{ChAH}, compare~\cite{ChGeroch,AshtekarLog,BeigSchmidt}) are absent in higher dimensions.

\bigskip
\noindent{\sc Acknowledgements:}
PTC was supported in part by
the Polish National Center of Science (NCN) under grant 2016/21/B/ST1/00940. PC is supported by St John's College, Cambridge through the Todd/Goddard Fund. Part of this research was performed while the authors were visiting the Institute for Pure and Applied Mathematics (IPAM) at UCLA, which is supported by the National Science Foundation (Grant No. DMS-1925919).

 \bibliographystyle{amsplain}
 
\bibliography{CameronChruscielIzero-minimal,%
../references/hip_bib,%
../references/reffile,%
../references/newbiblio,%
../references/newbiblio2,%
../references/chrusciel,%
../references/bibl,%
../references/howard,%
../references/bartnik,%
../references/myGR,%
../references/newbib,%
../references/Energy,%
../references/dp-BAMS,%
../references/prop2,%
../references/besse2,%
../references/netbiblio,%
../references/PDE}

\providecommand{\bysame}{\leavevmode\hbox to3em{\hrulefill}\thinspace}
\providecommand{\MR}{\relax\ifhmode\unskip\space\fi MR }
\providecommand{\MRhref}[2]{%
  \href{http://www.ams.org/mathscinet-getitem?mr=#1}{#2}
}
\providecommand{\href}[2]{#2}
\begin{thebibliography}{10}

\bibitem{AshtekarLog}
A.~Ashtekar, \emph{Logarithmic ambiguities in the description of spatial
  infinity}, Found.\ Phys. \textbf{15} (1985), 419--431. \MR{811916}

\bibitem{AshtekarHansen}
A.~Ashtekar and R.O. Hansen, \emph{A unified treatment of null and spatial
  infinity in general relativity. {I}. {U}niversal structure, asymptotic
  symmetries and conserved quantities at spatial infinity}, Jour.\ Math.\ Phys.
  \textbf{19} (1978), 1542--1566.

\bibitem{AshtekarRomano}
A.~Ashtekar and J.D. Romano, \emph{Spatial infinity as a boundary of
  spacetime}, Class.\ Quantum Grav. \textbf{9} (1992), 1069--1100. \MR{1158130}

\bibitem{ChBeig3}
R.~Beig and P.T. Chru\'{s}ciel, \emph{The asymptotics of stationary
  electro-vacuum metrics in odd spacetime dimensions}, Class.\ Quantum Grav.
  \textbf{24} (2007), 867--874, arXiv:gr-qc/0612012. \MR{MR2297271}

\bibitem{BeigSchmidt}
R.~Beig and B.G. Schmidt, \emph{Einstein's equations near spatial infinity},
  Commun.\ Math.\ Phys. \textbf{87} (1982/83), 65--80. \MR{680648}

\bibitem{BeigSimon2}
R.~Beig and W.~Simon, \emph{On the multipole expansion for stationary
  spacetimes}, Proc.\ Roy.\ Soc.\ London A \textbf{376} (1981), 333--341.

\bibitem{ChEnergy}
P.T. Chru\'{s}ciel, \emph{Lectures on energy in general relativity},
  \url{http://homepage.univie.ac.at/piotr.chrusciel/teaching/Energy/Energy.pdf}.

\bibitem{ChGeroch}
\bysame, \emph{On the structure of spatial infinity. {I}. {T}he {G}eroch
  structure}, Jour.\ Math.\ Phys. \textbf{30} (1989), 2090--2093. \MR{1009923}

\bibitem{ChAH}
\bysame, \emph{On the structure of spatial infinity. {II}. {G}eodesically
  regular {A}shtekar-{H}ansen structures}, Jour.\ Math.\ Phys. \textbf{30}
  (1989), 2094--2100. \MR{1009924}

\bibitem{ChDelay}
P.T. Chru\'{s}ciel and E.~Delay, \emph{On mapping properties of the general
  relativistic constraints operator in weighted function spaces, with
  applications}, M\'em.\ Soc.\ Math.\ de France. \textbf{94} (2003), 1--103
  (English), arXiv:gr-qc/0301073. \MR{MR2031583 (2005f:83008)}

\bibitem{ChGrant}
P.T. Chru\'{s}ciel and J.~Grant, \emph{{On Lorentzian causality with continuous
  metrics}}, Class.\ Quantum Grav. \textbf{29} (2012), 145001,
  arXiv:1110.0400v2 [gr-qc].

\bibitem{CorvinoSchoen2}
J.~Corvino and R.M. Schoen, \emph{On the asymptotics for the vacuum {E}instein
  constraint equations}, Jour.\ Diff.\ Geom. \textbf{73} (2006), 185--217,
  arXiv:gr-qc/0301071. \MR{MR2225517 (2007e:58044)}

\bibitem{Friedrich:i0}
H.~Friedrich, \emph{Gravitational fields near space-like and null infinity},
  Jour.\ Geom.\ Phys. \textbf{24} (1998), 83--163.

\bibitem{Gerochizero}
R.P. Geroch, \emph{{Structure of the gravitational field at spatial infinity}},
  Jour.\ Math.\ Phys. \textbf{13} (1972), 956--968. \MR{0309508}

\bibitem{Hansen}
R.O. Hansen, \emph{Multipole moments of stationary spacetimes}, Jour.\ Math.\
  Phys. \textbf{15} (1974), 46--52.

\bibitem{MinguzziLRR}
E.~Minguzzi, \emph{Lorentzian causality theory}, Living Rev.\ Rel. \textbf{22}
  (2019).

\bibitem{myersperry}
R.C. Myers and M.J. Perry, \emph{{Black holes in higher dimensional
  spacetimes}}, Ann.\ Phys. \textbf{172} (1986), 304--347.

\bibitem{Penrose:PRL63}
R.~Penrose, \emph{Asymptotic properties of fields and spacetimes}, Phys.\ Rev.\
  Lett. \textbf{10} (1963), 66--68. \MR{MR0149912 (26 \#7397)}

\bibitem{Prabhu}
K.~Prabhu, \emph{Conservation of asymptotic charges from past to future null
  infinity: supermomentum in general relativity}, Jour. High Energy Phys.
  (2019), 148, 47. \MR{3940857}

\end{thebibliography}

\end{document}